\documentclass[prl,superscriptaddress,amsmath,amssymb,showpacs,noshowkeys,a4paper,twocolumn]{revtex4-1}

\usepackage{graphicx}
\usepackage{dcolumn}
\usepackage{bm}
\usepackage[usenames]{color}
\usepackage{epsto	pdf}
\usepackage{wasysym}
\definecolor{mygray}{gray}{0.5}

\newcommand{\affpof}{\affiliation{Laboratoire de Physique et M\'ecanique des Milieux H\'et\'erog\`enes (PMMH), UMR
CNRS 7636 ; PSL - ESPCI, 10 rue Vauquelin, 75005 Paris, France; Sorbonne
Universit\'e - UPMC, Univ. Paris 06; Sorbonne Paris Cit\'e - UDD, Univ. Paris 07}}

\begin{document}

\title{Faraday wave lattice as an elastic metamaterial}

\author{L.~Domino}
\affpof
\author{M.~Tarpin}
\affpof
\author{S.~Patinet}
\affpof
\author{A.~Eddi}
\affpof

\begin{abstract}

Metamaterials enable the emergence of novel physical properties due to the existence of an underlying sub-wavelength structure. Here, we use the Faraday instability to shape the fluid-air interface with a regular pattern. This pattern undergoes an oscillating secondary instability and exhibits spontaneous vibrations that are analogous to transverse elastic waves. By locally forcing these waves, we fully characterize their dispersion relation and show that a Faraday pattern presents an effective shear elasticity. We propose a physical mechanism combining surface tension with the Faraday structured interface that quantitatively predicts the elastic wave phase speed, revealing that the liquid interface behaves as an elastic metamaterial.

%
%
\end{abstract}


\maketitle

%
%

%
%




An artificial material made of organized sub-wavelength functional building blocks is called a metamaterial \cite{Pendry1, MetaMat} when it exhibits properties that differ greatly from that of the unit cell. These new physical properties are intrinsic of the presence of an underlying structure. Although metamaterials are still strongly associated with negative index materials in optics \cite{NegIndexPendry}, they also refer to structures with mechanical \cite{Florijn:2014}, acoustic \cite{Lemoult:2013} or even thermodynamic properties \cite{Schittny:2013}. By engineering building blocks from micro to metric scale, several new mechanical properties emerge in metamaterials, such as cloaking in elastic plates \cite{CloakingElastic}, auxetic behavior \cite{Auxetics, PentamodeMetamaterials}, ultralight materials \cite{UltralightMeta} or seismic wave control \cite{SeismicCloaking}. So far the main challenge has been to design appropriate unit cells to obtain efficient metamaterial constructions. Here, we propose a novel approach that uses stationary waves to produce the underlying structure of a macroscopic metamaterial.

%

Spatial patterns arising in systems driven away from equilibrium have been extensively studied over the last decades \cite{Cross1993}. The Faraday instability is often used as a model system in non-linear physics and the patterns emerging from a vertically vibrated fluid layer are well documented \cite{Faraday1831, Ursell, FauveDouady, Douady, Kumar:1994}. This hydrodynamic instability appears at the interface between two fluids subjected to a vertical oscillation. Above a certain threshold of acceleration $a_c$, the surface shows a stationary deformation that oscillates at half the excitation frequency.
This pattern is both stable in time and regular in space, with a Faraday wavelength $\lambda_F$ defined by the inviscid gravity-capillary wave dispersion relation

\begin{equation}
\label{eq:1}
	\omega_F^2 = (gk_F + \frac{\sigma}{\rho}k_F^3 ) \tanh (k_Fh),
\end{equation}
where $k_F = 2\pi/\lambda_F$ is the Faraday wavenumber, $g=9.81$ m.s$^{-2}$ is the acceleration of gravity, $\sigma$ is the surface tension of the fluid, $h$ the fluid depth and $\rho$ its density.
For specific experimental conditions, one can achieve the formation of well structured and stable patterns (squares, hexagons, triangles... \cite{FauveDouady}). Although the pattern selection of this instability is quite complex, for a square vessel it is most often a square pattern that is obtained, with its two main directions aligned with the sides of the container. The pattern becomes unstable upon increasing the driving amplitude, and leads to a chaotic state \cite{Tufillaro} called ``defect-mediated turbulence" (DMT) \cite{DMT1989}. Transition to chaos is achieved by a phase instability called the oscillatory transition phase (OTP) \cite{FinebergDisorder, PhaseAnnulus, FauveDrift}. Similar oscillatory motions were observed and characterized in 1D systems such as Taylor-Couette \cite{TaylorCouette}, falling liquid columns \cite{LiqColumns}, Rayleigh-B\'enard convection rolls \cite{OscillatingRB}, and viscous fingering \cite{ImprimeurEckhaus}. Very few 2D systems exhibit this kind of secondary oscillatory modes: vibrated granular materials \cite{PhononGranular}, liquid columns \cite{LiqColumns2D} and bouncing droplets crystalline aggregates \cite{DropletCrystal}.

In this letter, we first show that the Faraday structure exhibits spontaneous in-plane transverse waves. We study their propagation in the 2D structure and link their existence to the emergence of an effective elastic shear modulus of the fluid-air interface. We propose a physical interpretation that quantifies the appearance of this effective mechanical property revealing that a Faraday wave lattice behaves as an elastic metamaterial.

\paragraph{Experimental set-up.~--} 

\begin{figure}
\includegraphics[width=0.94\linewidth]{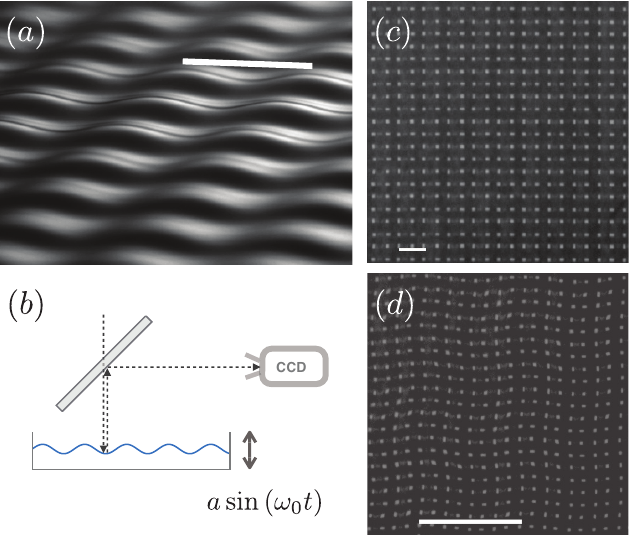} 
   \caption{
   (a) Side view of the standing Faraday instability wave pattern obtained for $\varepsilon>0$. The length of the white segment represents the Faraday wavelength (here $\lambda_F=5.1$ mm). (b) Sketch of the experimental setup. (c) Top view of the stable square pattern. The white segment has a length equal to $\lambda_F=5.1$ mm. (d) Top view of the oscillating Faraday pattern. The white segment has a length equal to $4\lambda_F=20.4$ mm, which is the wavelength of the spontaneous oscillations.}
   \label{fig:1}
\end{figure}

Our experimental set-up consists of a square vessel ($13$ cm$\times13$ cm) filled with a thin layer of silicone oil (viscosity $\eta=5~$mPa.s, density $\rho = 0.965$ kg.L$^{-1}$ and surface tension $\sigma = 20.9 $ mN.m$^{-1}$) of thickness $h=3$ to $5~$mm. The vessel is mounted on a vibration exciter (Br\" uel \& Kj\ae r), driven with a computer-controlled amplifier. The acceleration delivered by the vibration exciter is monitored using a calibrated accelerometer. The bath acceleration $a \cos{2\pi f_0 t}$ is sinusoidal, with frequency $f_0$ ranging from $72~$Hz to $120~$Hz. Above a given threshold acceleration $a_c$, the liquid interface spontaneously destabilizes and presents a regular square pattern of standing waves (see Fig. \ref{fig:1}(a)) with Faraday frequency $f_F=f_0/2$. The size of the pattern is about $25 \times 25$ Faraday wavelengths. We define the normalized control parameter $\varepsilon = (a - a_c)/a_c$.

The set-up and its imaging system is schematically shown in Fig. \ref{fig:1}(b). Diffused white light is shone on the container with a uniform square LED light and a beamsplitter inclined at $45^\circ$ enables us to image the vessel from the top, using a $2048 \times 2048$ pixels CCD camera. An example of the stable pattern obtained is shown in Fig. \ref{fig:1}(c), where only a few wavelengths are represented. This image is obtained by strobing the motion at an appropriate frequency, i.e. $18$ Hz when the forcing frequency $f_0$ is $72$ Hz (section \textit{Spontaneous secondary instability}) and $30$ Hz when the forcing frequency $f_0$ is $120$ Hz (section \textit{Forced vibrations}). Each white dot corresponds to an horizontal slope of the fluid interface, whether a maximum, a minimum or a saddle point \cite{Douady}. There are 4 white spots per Faraday unit cell [Fig. \ref{fig:1}(c)].

\paragraph{Spontaneous secondary instability.~--} 
\begin{figure}[b]
\includegraphics[width=1\linewidth]{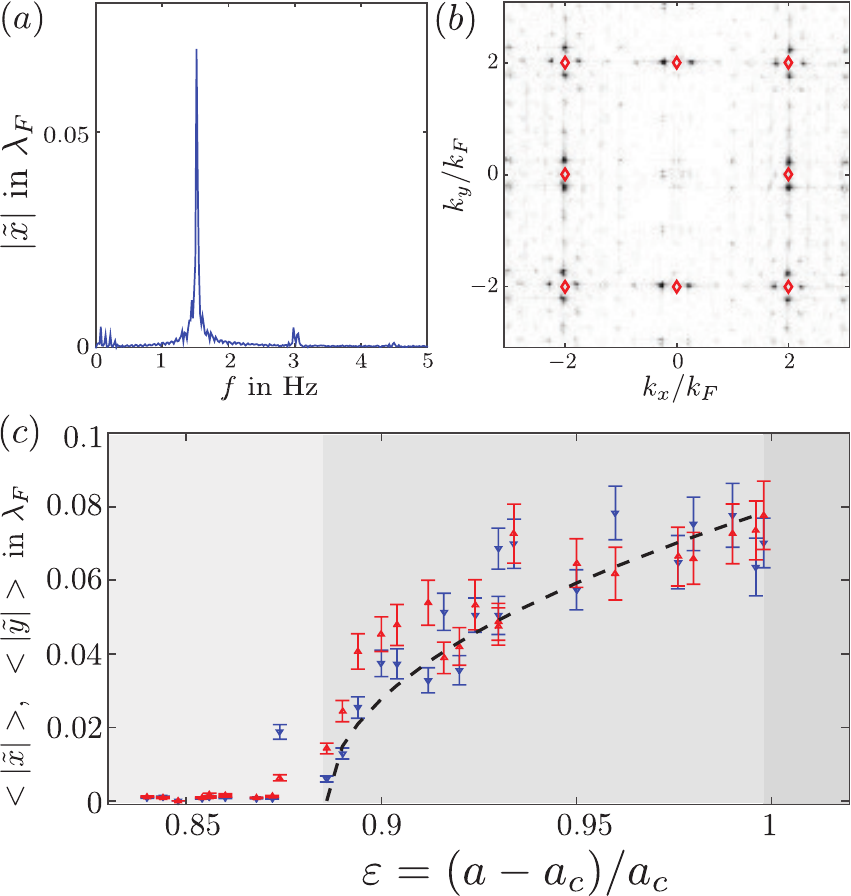}
  \caption{(a) Typical Fourier spectrum $\tilde{x}_{mn}(\omega)$ of the peak $(18,18)$ in the center of the pattern. 
  (b) Modulus of spatial Fourier 2D spectrum $|\hat{y}(k_x, k_y, f)|$ for $\varepsilon = 0.976$, averaged for $f=1.57 \pm 0.1$ Hz. Open diamonds show the position of the Fourier peaks for a stable pattern [Fig. \ref{fig:1}(c)]. 
  (c) Amplitude of the spontaneous vibration as a function of the normalized control parameter, averaged over all the antinodes of the lattice. 
  The up (red) and down (blue) triangles correspond to $|\tilde{y}|$ and $|\tilde{x}|$, respectively. 
  Dashed line is a square root fit. 
  Grey shades denote (from left to right): stable pattern, spontaneous vibrations of the lattice and chaotic behavior.}
\label{fig:2}
\end{figure}

Upon increasing the driving amplitude to about twice the threshold value, spontaneous oscillations of the square lattice appear (Fig. \ref{fig:1}(d) and Supplementary Movies 1 and 2). These oscillations are in-plane modulations of the pattern along its two main directions. They exhibit a spatial periodicity $\lambda = 4\lambda_F$, corresponding to the white segment presented in this figure. We label each bright spot with indexes $(m,n)$ and we detect their in-plane position $(x_{mn}(t),y_{mn}(t))$ using a standard custom Matlab algorithm. 
A typical spectrum corresponding to the parameters of Fig. \ref{fig:1}(d) is presented in Fig. \ref{fig:2}(a). The measured frequency is $f=1.52~$Hz with an amplitude of $0.07~\lambda_F$. 
To analyse in more detail the spatial structure of the lattice dynamics, we perform spatio-temporal Fourier transforms. In Fig. \ref{fig:2}(b) we show a typical spatial 2D spectrum obtained at $1.57 \pm 0.1$ Hz. The Fourier peaks corresponding to the stable Faraday pattern (open diamonds) are split into two symmetric sub-peaks indicating the standing nature of the pattern oscillations. These sub-peaks are located at a distance $k_f/4$ from the original one, confirming the wavelength observed in Fig. \ref{fig:1}(d). The peak in the $k_x$ direction (resp. $k_y$) is split in the $k_y$ direction (resp. $k_x$) revealing that the spontaneous oscillations correspond to the propagation of a standing transverse wave in the initial Faraday square lattice. These spontaneous oscillations of the pattern correspond to a Hopf bifurcation that takes place close to the threshold of transition to chaos. This is confirmed when measuring their amplitude at the vibration frequency as a function of the control parameter $\varepsilon$, where we observe the supercritical nature of this bifurcation [Fig. \ref{fig:2}(c)]. This figure also shows that the amplitude of the vibrations is the same for both directions. At $\varepsilon \simeq 1$, the pattern becomes unstable and we observe the formation of defects. We note that this threshold value is different from what was found elsewhere \cite{FinebergDisorder, Tufillaro} as it depends on the depth of the liquid layer, the fluid viscosity and the forcing frequency. 

%
%
%

%
%
%
%
%
%

Here, we want to point out that the spontaneous oscillations occur at a frequency $f$ much lower than the Faraday frequency $f_F$ whereas their spatial wavelength $\lambda\simeq2.0~$cm is $4$ times larger than $\lambda_F$. In our experimental conditions and at this frequency $f$, the gravito-capillary dispersion relation (eq. 1) gives a wavelength of $\lambda_{gc}=23.26~$cm much larger than $\lambda$. This means that the transverse standing wave responsible for the pattern oscillations is governed by a different physical mechanism.



\paragraph{Forced vibrations.~--}

\begin{figure}[t]
\includegraphics[width=1\linewidth]{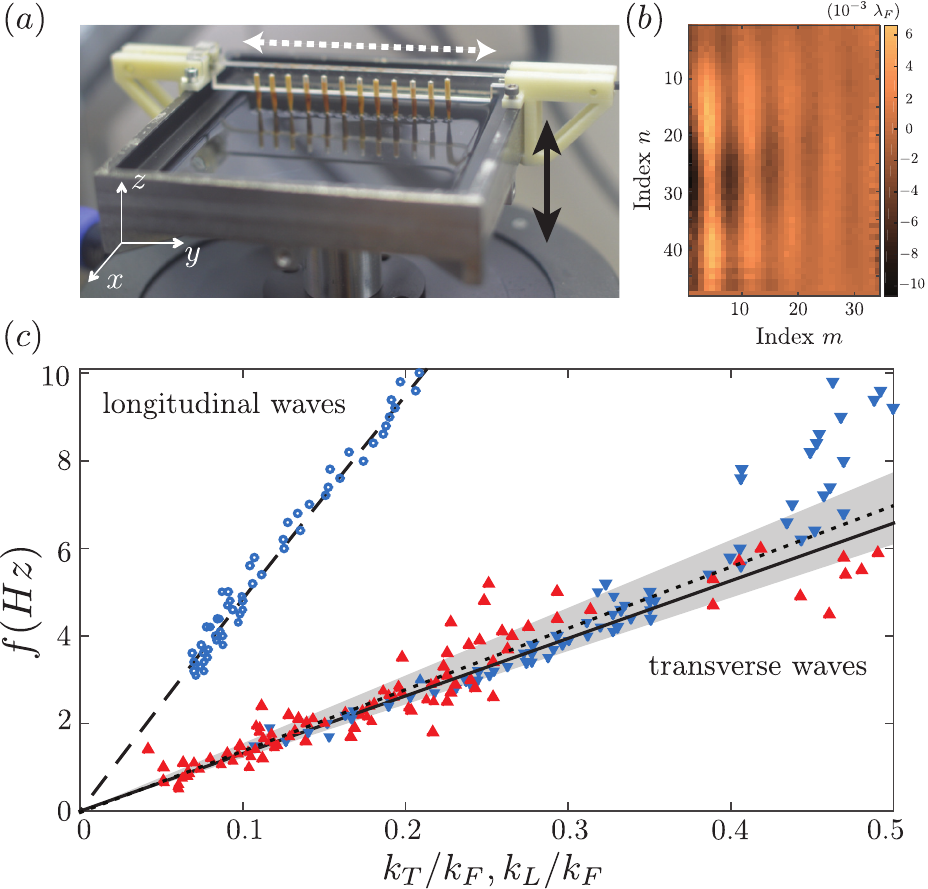}
    \caption{(a) Device used to force the vibrations of the Faraday pattern. 
    The black arrow shows the vertical motion of the whole vessel, the dotted white arrow shows the direction of the comb vibration. The width of the vessel is 12 cm. 
    (b) Map of the real part of the FFT peak $Re (\tilde{y})$ for a forcing frequency of $3.7${Hz}. The forcing device is on the left, each pixel represents a bright point of our images. 
    (c) Dispersion relation $f(k_T)$ and $f(k_L)$. 
    Blue down triangles: transversal waves in the $y$ direction. 
    Red up triangles: transversal waves in the $x$ direction. 	
    Open circles: longitudinal waves in the $y$ direction.
    Solid black line: linear fit.
    Dashed line: gravity-capillary wave dispersion relation.
    Dotted line and grey background: prediction from eq. \ref{prediction} and its associated uncertainty.}
    \label{fig:3}
  \end{figure}

We now investigate the characteristics of these oscillating modes of the Faraday wave pattern by forcing the vibrations of stable square patterns. We set the Faraday vertical forcing frequency to $f_0=120~$Hz (resulting in $\lambda_F=3.5~$mm), the liquid depth to $h=3~$mm and the forcing acceleration to $\varepsilon=0.81$ in order to get a stable and larger initial Faraday square pattern (its size is now $35 \times 35$ Faraday wavelengths). We add to the vessel a custom-made forcing device consisting of a comb dipping into the liquid to a small depth [Fig. \ref{fig:3}(a)]. It is mounted so that it is aligned with one side of the container, and it vibrates vertically along with it. The comb is set in motion by a second vibration exciter (Br\" uel \& Kj\ae r) to oscillate horizontally in the reference frame of the container at frequencies ranging from $0.5${ Hz} to $10${ Hz}. The distance between the comb teeth is set to $2 \lambda_F$, and the amplitude of the forcing sinusoidal motion is set to half the Faraday wavelength. This allows us to generate a sinusoidal oscillation of the line of Faraday peaks located below the forcing comb.

We observe a transversal wave that propagates away from the forcing device at the forcing frequency $f$. We detect the position $(x_{mn}(t),y_{mn}(t))$ of each bright spot and perform a temporal Fourier transform to obtain $(\tilde{ x}_{mn}(f),\tilde{y}_{mn}(f))$. Fig. \ref{fig:3}(b) displays $Re \left (\tilde y(f) \right )$ for the excitation frequency $f=3.7$ Hz. We observe a periodic pattern that decays along the $x$ direction (indexed as $m$) away from the forcing device. This corresponds to the propagation of a transverse wave in the $x$ direction at the forcing frequency $f$ (the motion is along $y$).
We define the spatial phase $\phi_y$ of the pattern vibration along $y$ as $\phi_y(x,t) = \phi_0 \exp[(-\alpha + ik_T) x + i2\pi f t]$ where $1/\alpha$ is the decay length of the oscillation and $k_T$ its wavenumber. From the experimental data we extract $\alpha$ and $k_T$ for each value of $f$. The decay length $1/\alpha$ does not depend significantly on $f$ and its typical value is $1/\alpha\simeq11.5\lambda_F$, whereas the value of $k_T$ depends on $f$. Due to the imperfections of the forcing device, we also notice the presence of a periodicity in the $y$ direction (indexed as $n$), corresponding to a longitudinal wave propagating in the $y$ direction with wavenumber $k_L$ that we extract form Fig. \ref{fig:3}(b). We perform the same analysis on $Re(\tilde x)$, for which we have similar maps as Fig. \ref{fig:3}(b). 
Altogether, we report the existence of transverse waves along both the $x$ and $y$ direction, as well as longitudinal waves. Fig. \ref{fig:3}(c) presents the dispersion relations $f(k_T)$ (blue and red triangles) and $f(k_L)$ (open circles) that we obtain for $f$ ranging from $0.5$ to $10$ Hz. 
We first notice that $f(k_L)$ obeys the standard surface waves dispersion relation predicted by eq. 1 (dashed line). This means that the forcing device induces gravito-capillary waves. Their dispersion relation appears quite linear in Fig. \ref{fig:3}(c) since the shallow water approximation applies ($k_Fh \ll 1$). On the other hand, the dispersion relation for transverse waves $f(k_T)$ is markedly different. We observe a linear increase of $f$ with $k_T$ with a much lower slope. A linear fit gives the phase speed of the transverse waves $c_T=4.60~$cm.s$^{-1}$.

\paragraph{Physical interpretation.~--}
\label{parMechanism}

These experimental results show that there exists a new type of wave propagating at the fluid-air interface. 
They are transverse waves associated to the presence of a pre-existing Faraday wave pattern and reminiscent of 2D shear waves that propagate in elastic media. 
Here we present a quasi-2D model in which we identify the Faraday cellular pattern to a 2D metamaterial with solid-like properties. Indeed, transverse waves in an elastic material propagate with constant phase velocity $c_T$ that only depends on the elastic shear modulus $\mu$. We use the structure of the Faraday wave lattice and the fluid properties to derive an effective elastic shear modulus and quantitatively predict the transverse waves properties.

\begin{figure}
\includegraphics[width=1\linewidth]{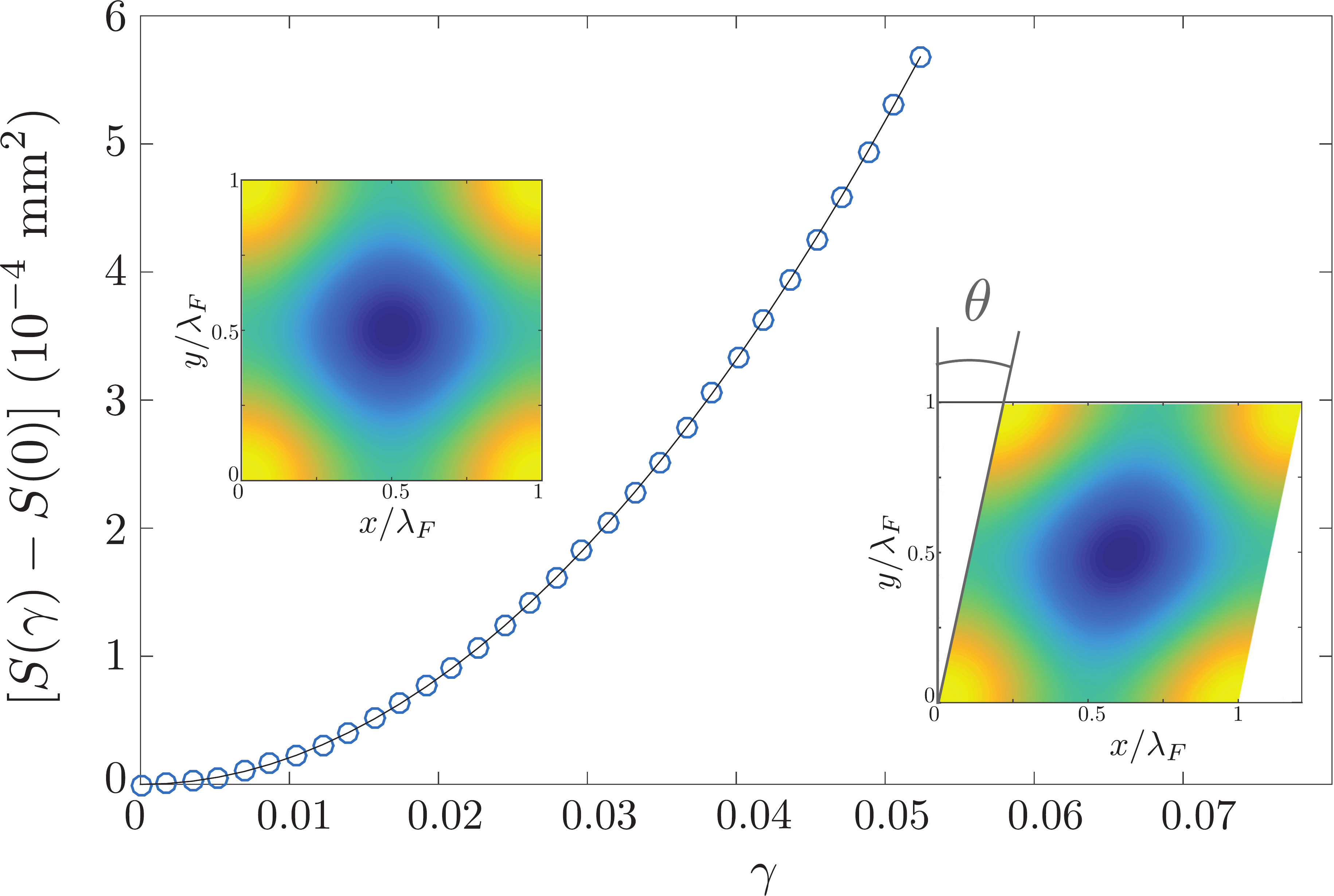} 
    \caption{Open circles: evolution of $S(\gamma)$ with $\gamma$ computed numerically for $\lambda_F = 3.5$mm and $A_0/\lambda_F = 13.5\%$. Dark line: theoretical prediction from eq. \ref{eq:surf_gamma}.
    Inset (left): Reference surface. 
    Inset (right): Sheared surface, with $\tan{\theta} = \gamma$.
    }
    \label{fig:4}
  \end{figure}

We consider a reference state for the interface defined as 
\begin{equation}
		z_0 (x,y,t) = A(t) \cos \left (\pi \frac{ x+y}{\lambda_F } \right ) \cos \left (\pi \frac{ x-y}{\lambda_F }  \right ),\\
\label{eq:2}
\end{equation}
with $A(t)=A_0\cos (2 \pi f_F t)$ the amplitude of the stationary wave. Measurements (e.g. on Fig. \ref{fig:1}(a)) give $A_0/ \lambda_F =13.5\% \pm 3.5\%$. This 2D function gives a succession of peaks and crests arranged in a square pattern tilted at $45 ^\circ$, as represented in Fig. \ref{fig:4} (inset on the left). We apply a shear strain $\gamma = \tan {\theta}$ to this elementary cell [Fig. \ref{fig:4} (inset on the right)] and calculate analytically its surface area $S(\gamma)$
%
%
%
%
%


\begin{equation}
\label{eq:surf_gamma}
	S(\gamma)  =    \displaystyle\int\limits_{0}^{\lambda_F} \mathrm{d}y \displaystyle\int\limits_{\gamma y}^{\lambda_F+\gamma y}  \mathrm{d}x 
	\left  [1 + \left ( \frac{\partial z_0}{\partial x} \right ) ^2 + \left (  \frac{\partial z_0}{\partial y}  - \gamma  \frac{\partial z_0}{\partial x} \right ) ^2 \right ]^{1/2} \\
\end{equation}
As $f \ll f_F$ we average $S(\gamma)$ in time which corresponds to replacing $A(t)$ with $\cal{A}$$=A_0 2/\pi $. Fig. \ref{fig:4} shows the numerical evaluation of $S(\gamma)$ for $A_0/\lambda_F = 13.5\%$.
  As $S(\gamma)$ is an even function ($\gamma$ and $-\gamma$ give the same area),  $\left . \frac{\partial S}{\partial \gamma} \right |_{\gamma = 0} = 0$. For a non-zero amplitude of the Faraday wave, the shearing deformation leads to a surface excess $\Delta S=S (\gamma) - S(0)$ that we can approximate for small deformations
\begin{equation*}
\begin{split}
\Delta S & = \left. \frac{1}{2} \frac{\partial ^2S}{\partial \gamma^2} \right| _{\gamma = 0} \gamma^2 = \frac{1}{2} S_{\gamma \gamma} \gamma^2. \\
 	\end{split}
\end{equation*}

For our experimental parameters ($\lambda_F = 3.5$mm and $A_0/\lambda_F = 13.5\%$) we obtain $S_{\gamma\gamma} \simeq 4.13 ~10^{-7}$ mm$^2$.

Due to surface tension there is an energy cost that depends on the applied shear deformation $\Delta E(\gamma)=\sigma \Delta S (\gamma)$.  We then define the effective elastic energy density per unit area $W_S = \sigma \Delta S/\lambda_F^2$ (in J.m$^{-2}$) and introduce the effective shear modulus $\mu_S$ of the Faraday wave pattern: $W_S= 2 \mu_S \epsilon_{xy}^2$ where $\epsilon_{xy} = \frac{1}{2} \gamma$. 
Following standard elasticity theory \cite{LIFSHITZ:1986qf} the transverse elastic wave phase velocity $c_T$ in a 2D elastic medium is written

\begin{equation}
	 c_T = \sqrt{\frac{\mu_S}{\rho_S}} = \sqrt{\frac{\sigma S_{\gamma \gamma}} {\rho_S \lambda_F^2}} ~ , \\
\label{prediction}
\end{equation}
with $\rho_S$ the density per unit area, defined as $\rho_S = \rho{\cal A}$. 

Using equations \ref{eq:2} and \ref{eq:surf_gamma}, the velocity we obtain is $c_T= 4. 84 \pm 0.63$ cm.s$^{-1}$, which is in excellent agreement with the experimental result of $4.60$ cm.s$^{-1} $. We represent in Fig. \ref{fig:3}(c) the estimated dispersion relation (dotted line), the grey background representing the uncertainty. 
 
\paragraph{Conclusion.~--}

We have characterized a new secondary instability that arises in 2D Faraday patterns close to the transition towards chaos. This instability leads to vibrations of the Faraday pattern similar to a 2D transverse elastic wave. We established the dispersion relation for these waves and showed that it differs markedly from the standard gravito-capillary waves that propagate at the liquid-air interface. We propose a physical mechanism that combines the surface tension with the pre-existing Faraday wave structure at the interface. We are able to derive an effective shear modulus $\mu_S$ for the Faraday wave pattern that quantitatively agrees with the experimental observations. 

In this work, we observe the emergence of a new physical property, namely an effective 2D elasticity, at the liquid-air interface. Our interpretation reveals that it is intimately related to the existence of a periodic pattern imprinted on the liquid interface. From this perspective, the Faraday wave pattern creates a mechanical metamaterial at macroscopic scale. In the future, we would like to investigate in more detail the limit $k/k_F=1/2$ corresponding to the edge of the first Brillouin zone in a crystalline material. Another line of future research is to understand if there exists a second elastic constant for the medium as in usual elastic solids. More generally, wave-based metamaterials offer unique possibilities as wavelengths and patterns can be dynamically tuned.

%
%
%
%
%


\acknowledgments{The authors would like to thanks E. Fort and Y. Couder for fruitful discussions as well as X. Benoit-Gonin and A. Fourgeaud for their help in setting up the experiments.}

%



\end{document}